\documentclass{vldb}
\usepackage{graphicx}
\usepackage{balance}  
\usepackage{listings}
\usepackage{paralist}
\usepackage[justification = centering]{caption}
\usepackage{booktabs}
\usepackage{float}
\usepackage{comment}
\usepackage{dirtytalk}

\newdef{definition}{Definition}
\newdef{lemma}{Lemma}

\vldbTitle{Extending Databases to Support Data Manipulation with Functional Dependencies: a Vision Paper}
\vldbAuthors{Nikita Bobrov, Kirill Smirnov, George Chernishev}
\vldbDOI{https://doi.org/10.14778/xxxxxxx.xxxxxxx}
\vldbVolume{xxx}
\vldbNumber{xxx}
\vldbYear{xxx}

\begin{document}


\title{Extending Databases to Support Data Manipulation with Functional Dependencies: a Vision Paper}



%
%
%
%

\numberofauthors{3} 

\author{
%
%
\alignauthor
Nikita Bobrov\\
        \affaddr{JetBrains Research}\\
        \affaddr{Saint-Petersburg, Russia}\\
        \email{nikita.v.bobrov@gmail.com}
%
%
\and
\alignauthor
Kirill Smirnov\\
        \affaddr{JetBrains Research}\\
        \affaddr{Saint-Petersburg, Russia}\\
        \email{kirill.k.smirnov@gmail.com}
\alignauthor
George Chernishev\\
        \affaddr{JetBrains Research}\\
        \affaddr{Saint-Petersburg, Russia}\\
        \email{chernishev@gmail.com}
}
\date{30 July 1999}


\maketitle

\begin{abstract}

 In the current paper, we propose to fuse together stored data (tables) and their functional dependencies (FDs) inside a DBMS. We aim to make FDs first-class citizens: objects which can be queried and used to query data. Our idea is to allow analysts to explore both data and functional dependencies using the database interface. For example, an analyst may be interested in such tasks as: ``find all rows which prevent a given functional dependency from holding'', ``for a given table, find all functional dependencies that involve a given attribute'', ``project all attributes that functionally determine a specified attribute''.

For this purpose, we propose: 
\begin{inparaenum}[(1)]
\item an SQL-based query language for querying a collection of functional dependencies
\item an extension of the SQL SELECT clause for supporting FD-based predicates, including approximate ones
\item a special data structure intended for containing mined FDs and acting as a mediator between user queries and underlying data.
\end{inparaenum} We describe the proposed extensions, demonstrate their use-cases, and finally, discuss implementation details and their impact on query processing.

\end{abstract}

\section{Introduction}


The concept of a functional dependency (FD) was proposed more than $45$ years ago. Initially, it was employed by data administrators for schema normalization. In this use-case, administrators were aware of existing FDs due to their domain knowledge. Nowadays, FDs are also used in a different task: discovering all FDs that are contained in an unfamiliar dataset and presenting them for analysis.

In the recent years, the mining of functional (both exact and approximate), conditional, inclusion, and other types of dependencies has experienced a surge of interest~\cite{Caruccio:2017:EMR:3102254.3102259, Papenbrock:2016:HAF:2882903.2915203, Shaabani:2018:IEI:3269206.3271724}. Among others, the Metanome project~\cite{Metanome} offers a plethora of high performance algorithms for mining all kinds of dependencies.

So far, mined functional dependencies have existed as relatively passive database objects. Of course, they were used for data cleaning and analysis~\cite{Fan:2008:CFD:1366102.1366103, Pyle:1999:DPD:299577, UGuide}, as well as in several proposals for query optimization~\cite{Eich:2016:FPG:2977797.2977802, paulley2001exploiting}. However, FDs still were external objects, and DBMSes were mostly unaware of them.

Our idea is to ``attach'' functional dependencies to their respective table and facilitate their manipulation, including:
\begin{enumerate}
    \item capabilities to query functional dependencies contained in a table;
    \item capabilities to impose various filtering conditions using functional dependencies while querying the data itself.
\end{enumerate} 

The first allows to easily navigate a collection of mined functional dependencies. Despite the fact that only minimal, non-trivial functional dependencies\footnote{Further on in this paper, we assume that all mentioned FDs are minimal and non-trivial.}~\cite{DynFD, Papenbrock:2016:HAF:2882903.2915203} are of interest to users, presenting all of them for analysis is still a problem: a table with several dozens of attributes may have millions of FDs. 

State-of-the-art approaches output FDs either as a plain list or using simple visualization techniques~\cite{Metanome}, which are not suitable even for medium-sized tables. Therefore, we propose a special language which will allow to declaratively query a collection of FDs, making data navigation easy. 

A couple of examples of user needs (and queries): ``for a given table, find all FDs that involve a given attribute'', ``for a given table and a given attribute find all FDs that functionally determine it''.


The second allows to perform in-depth analysis of data. For example, an analyst may be interested in such queries as: ``find all rows which prevent a given functional dependency from holding'' or ``project all attributes that functionally determine a specified attribute''.

Such analysis would be of use for business and scientific applications, where data is represented by wide tables. For example, the Sloan Digital Sky Survey (SDSS) dataset contains tables that feature more than $400$ attributes~\cite{SDSS}. 

It is worthwhile to move such analysis inside a database due to the following:
\begin{enumerate}
    \item SQL and RDBMes are immensely powerful and convenient tools to query and explore data. These qualities stem from their declarative nature and the presence of a query optimizer, which give unparalleled flexibility. Such capabilities are needed during the FD-related data exploration. For example, one may want to re-check the presence of a FD on a subset of a table (see example in Listing~\ref{listing:filtering}). Filtering it outside is inconvenient and costly.
    \item FDs are inseparable from data in a sense that they should be stored together with the data they belong to. Moving them outside will require synchronization in case of changes in data. This can be prohibitively expensive in some cases.
    \item Finally, there is a trend for in-database analytic processing~\cite{AboKhamis:2018:ILS:3196959.3196960, Hellerstein:2012:MAL:2367502.2367510, Feng:2012:TUA:2213836.2213874}. Note that similar integration happened in XML processing, temporal extensions, and is currently happening in machine learning~\cite{Kumar:2016:JJT:2882903.2882952, Schleich:2016:LLR:2882903.2882939, Kumar:2015:LGL:2723372.2723713}.
\end{enumerate}

In this paper we continue to develop our proposal~\cite{DBLP:conf/cidr/Chernishev20} for extending RDBMSes with capabilities to manipulate data using FDs. In the current work we present five motivating examples, an intended workflow for such system, and our initial view on how to implement it. Finally, we present emerging problems and challenges.



Overall, the contribution of this paper is:
\begin{enumerate}
    \item A draft of two query languages, one for manipulating a collection of functional dependencies, and another for using FD-related conditions inside the SELECT clause.
    \item A proposal of a storage mechanism for FDs along with a special FD mining operator.
    \item A discussion of problems, challenges, and research venues of query processing techniques for handling both data and FDs.
\end{enumerate}

\section{Background}

\subsection{Functional Dependencies: Basics}
\label{sec:Basics}

 
In this section, we describe concepts that are necessary for understanding the FD-related part of our proposal. We follow the notation of the PYRO~\cite{Pyro} paper. 

An $exact$ functional dependency is defined as follows:
\begin{definition}
Given a relational schema $R$ and an instance $r$ over $R$ with attribute sets $X, Y \subset R$, we say that a functional dependency $X \rightarrow Y$ holds iff for any $t_1, \
 t_2 \in r$, the following is true: if $t_1[X] = t_2[X]$, then $t_1[Y] = t_2[Y]$.
Henceforth, we call the determinant set of attributes $X$ the left-hand side (LHS) of the FD, and the dependent set the right-hand side (RHS). 
\end{definition}

Moreover, we consider the class of $approximate$ functional dependencies (AFDs), which features an error threshold $e_{max}$. The idea is that a specified value $e_{max}$ defines the fraction of tuples or pairs of tuples which can violate the FD.


\begin{definition}
Given an instance $r$ and an AFD candidate $X \rightarrow Y$, its error is calculated as: 
$$e(X \rightarrow Y, r) = \frac{|\{(t_1, t_2) \in r^2 | t_1[X] = t_2[X] \wedge t_1[Y] \neq t_2[Y]\}|}{|r|^2 - |r|}$$
AFD $X \rightarrow Y$ holds on $r$ if $e(X \rightarrow Y, r) \leq e_{max}$.
\label{def:ErrorMeasure}
\end{definition}

An example demonstrating the calculation of error is presented further, see Listing~\ref{ex:error}.

However, only a small number of dependencies is usually known in a database. Discovering them was a duty of a data administrator, and it is actually a part of the integrity constraint design problem. Still, in an overwhelming number of cases no dependencies are known for a given database. This leads to the problem of database dependency discovery that has received a great deal of attention since the 90's. A significant number of mining techniques and, consequently, algorithms were developed to tackle the problem of both exact ~\cite{TANE, Fdep, Papenbrock:2016:HAF:2882903.2915203} and approximate~\cite{AIDFD, Pyro} FD discovery. The main purpose of these algorithms is to provide a set of non-trivial and minimal dependencies (also called the $gold \  standard$~\cite{AIDFD}, $canonical$, $irreducible$ set).

\begin{definition}
An exact or approximate FD $X \rightarrow Y$ is called:
\begin{inparaitem}
 \item[(a)] trivial, if $Y \subset X$;
 \item[(b)] minimal, if $Y$ is not functionally dependent on any subset of $X$.
\end{inparaitem}
\end{definition}


Since our proposal also considers the specification of \textit{conditions} on tuple values while querying with FD-based predicates, we need to recall the definition of a conditional functional dependency ~\cite{bohannon_conditional_2007}:

\begin{definition}
A CFD $\phi$ on $R$ is a pair $(X \rightarrow Y, T_p)$, where $X \rightarrow Y$ is a FD, and $T_p$ is a\ ``pattern tableau'' that defines over which rows of the table the FD $X \rightarrow Y$ should hold. Each entry $t_p \in T_p$ specifies a pattern over $X \cup Y$, so for each attribute in $A \in X \cup Y$, either $t_p[A] = \alpha$, where $\alpha$ is a value in the domain of $A$, or a special wildcard symbol $t_p[A]=\underline{\quad}$. A row $r_i \in R$ satisfies an entry $t_p$ of tableau $T_p$ for attributes $A$ ($r_i[A] \asymp t_p[A]$), if either $r_i[A] = t_p[A]$ or $t_p[A]=\underline{\quad}$. The CFD $\phi$ holds, if: \\
\centerline{$\forall i,j,p.r_i[X] = r_j[X] \asymp
        t_p[X] \Rightarrow r_i[Y] = r_j[Y] \asymp t_p[Y]$.}
\end{definition}

As we will show, the concept of CFDs quite naturally arises when the classic SQL WHERE clause is specified with conditions along with FD-based predicates. An example Table~\ref{table:CFDExample} demonstrates how CFDs allow to unambiguously describe subsets of rows which should hold a certain FD. The pattern tableau $T_1$ should be interpreted in the following way:
\begin{itemize}
\item if two tuples $t_1, t_2$ agree on $STR$ and $t_1[CC] = t_2[CC] = 01$, $t_1[AC] = t_2[AC] = 908$, then they should agree on $ZIP$ and $t_1[CT] = t_2[CT] = MH$;
\item if two tuples $t_1, t_2$ agree on $STR$ and $t_1[CC] = t_2[CC] = 01$, $t_1[AC] = t_2[AC] = 212$, then they should agree on $ZIP$ and $t_1[CT] = t_2[CT] = NYC$;
\item any other tuples with no specified conditions on values (i.e., wildcard symbol) are treated as regular tuples that should agree on arbitrary values.
\end{itemize}

\begin{table}[H]
    \begin{minipage}{0.5\textwidth}
    \centering
    \caption{CFD $\varphi_2 = ([CC, AC, STR] \rightarrow [CT, ZIP], T_1)$}
    \begin{tabular}{|l|l|l|l|l|l|}
        \hline
        \textbf{CC} & \textbf{AC} &  \textbf{NM} & \textbf{STR} & \textbf{CT} & \textbf{ZIP}\\ \hline
        01  & 908  & Mike & Tree Ave.& MH & 07974 \\ \hline
        01  & 908  & Rick & Tree Ave.& MH & 07974 \\ \hline
        01  & 212  & Joe  & Elm Str. & NYC & 01202 \\ \hline
        01  & 212  & Jim  & Elm Str. & NYC & 02404 \\ \hline
        01  & 215  & Ben  & Oak Ave. & PHI & 02394 \\ \hline
        44  & 131  & Ian  & High St. & EDI & EH4 1DT \\ \hline
    \end{tabular}
    \label{table:CFDExample}
    \end{minipage}
    %
    
    \vspace{0.5cm}
    \begin{minipage}{0.5\textwidth}
    \centering
    \caption*{Tableau $T_1$}
    \begin{tabular}{|l|l|l||l|l|}
        \hline
        \textbf{CC} & \textbf{AC}  & \textbf{STR} & \textbf{CT} & \textbf{ZIP}\\ \hline
        --- & --- &  --- & --- & ---\\ \hline
        01 & 908 &  --- & MH & ---\\ \hline
        01 & 212 & --- & NYC & ---\\ \hline
    \end{tabular}
    \label{table:CFDTableau}
    \end{minipage}
\end{table}

Regular CFDs allow to specify values of tableau patterns as single constants, but as it was shown in~\cite{bhowmick_analyses_2009}, conditions can also be described with operators $\neq, <,>,\leqslant, \geqslant$. Such an extension comes very handy, since specification of ranges is a common requirement to SQL-like languages. Now, with the expressiveness of CFDs, we are able to use WHERE-like clauses inside FD-based predicates. 

An approximate CFD can be defined with the two following metrics:
\begin{definition}
The support of CFD $\phi = (X \rightarrow A, t_p)$ on relation $r$, denoted $sup(\phi, r)$, is defined by:
	$$sup(\phi, r) = \frac{|r_{t_p}|}{|r|},$$ where $|r_{t_p}| = |r_{t_p[XA]}|$ is the number of tuples in $r$ matching $t_p$ on a set of attributes $XA$.
\end{definition}

\begin{definition}
The confidence of CFD $\phi = (X \rightarrow A, t_p)$ on relation $r$, denoted $conf(\phi, r)$, is defined as
$$conf(\phi, r) = \frac{max\{|r'|, r' \subseteq r, r' \vDash (X \rightarrow A, t_p)\}}{|r|}$$
\end{definition}

CFD discovery differs considerably from FD discovery. First, algorithms are classified by the type of discovered CFDs: constant (the pattern tableau does not contain the wildcard symbol) and general CFD (any symbol or constant allowed). Second, the process of discovering CFDs is more complicated by its nature~--- not only a FD must be validated, but all possible pattern tableaus for which this FD is satisfied must be found. Thus, it is possible that for a given FD a number of CFDs with different tableaus may exist, which explains why the minimal set of CFDs is much larger than the set of all minimal FDs.
Third, CFDs are very similar to association rules. Moreover, the association rule $(X, t_p) \Rightarrow (A, a)$ with confidence = 1, is actually a constant CFD $\varphi = (X \rightarrow A, (t_p || a))$, where:
$t_p$~--- pattern tableau each row of which is a constant of the domain $X$, and $a$ is a single value of the domain $A$.


\subsection{Mining Functional Dependencies:~Modern Data Structures and Algorithms}


The core data structure in the modern FD discovery process is Position List Index. PLI is based on a concept of partition (denoted by $\pi$), which allows to represent groups of attributes in such a way that the process of verifying FDs becomes a special case of intersection of two sets. 
PLI is comprised of sorted clusters, which are basically lists of tuple indices that agree on certain values for a group of attributes X.

\begin{definition}
	Let $r$ be a relation with schema $R$, and let $X \subseteq R$ be a set of attributes. A \textit{cluster} is a set of all tuple indices in $r$ that have the same value for $X$, i.e., $c(t) = \{i \vert t_i \left[ X \right] = t \left[ X \right] \}$. The PLI of $X$ is the set of all such clusters except for singleton clusters:
	$$\overline{\pi}(X) := \{c(t) \vert t \in r \wedge \left\vert c(t) \right\vert > 1 \}. $$
\end{definition}

The process of FD validation over PLIs is performed according to a lemma:

\begin{lemma}
The FD $X \rightarrow Y$ holds, iff $|\bar{\pi}(X)| = |\bar{\pi}(X\cup Y)|$.
\end{lemma}

As we can see, the partition concept plays a vital role in the functional dependency discovery theory: a special class of algorithms that employ the lemma exists. Such algorithms organize the discovery process as lattice traversal. The search space is represented by a lattice which contains attribute combinations that need to be validated. The validation itself is a rather costly operation of partition intersection that is performed to check lemma conditions. Since the discovery problem is exponential in the number of attributes, lattice traversal algorithms fall short in performance as a dataset becomes ``wider''.

Other extensively used data structures are agree- and dif\-fe\-ren\-ce- sets. An agree-set is a group of attributes that agree on the values in certain tuple pairs, while a difference-set is a complement to an agree-set and is used to infer valid FDs. Discovery algorithms need to perform pair-wise record comparison in order to derive agree-sets, which makes the discovery problem quadratic in the number of rows. Despite the fact that approach solely based on agree-sets fails on ``long'' datasets, the data structure itself is still used in modern algorithms\cite{Pyro, Papenbrock:2016:HAF:2882903.2915203} when performing sampling-based FD discovery and constructing positive and negative covers.

A comprehensive study on a number of existing algorithms can be found in~\cite{Papenbrock:2015:FDD:2794367.2794377}. 


\section{Motivation}

In this section we discuss several use-cases that can be addressed using the proposed framework. These are mostly ad-hoc scenarios that involve in-database data  exploration and repairs.



\subsection{Data repair preparation and post-repair validation}~\label{sec:Motivation:DRP}

To the best of our knowledge, the majority of data cleaning tools that employ dependencies accept a precomputed set of FDs or CFDs as an input~\cite{DCClean:Papotti:2013, NADEEF:Dallachiesa:2013, BigDancing:Khayyat:2015}. Therefore, it should be somehow obtained and there are only two possible sources of dependencies:
\begin{enumerate}
    \item automatic tools that mine dependencies (e.g. Metanome);
    \item domain knowledge of the user.
\end{enumerate}

Using the output of mining tools for performing repairs is not viable, since they produce a large and incomprehensible list of dependencies. At the same time, the user is interested in only a small fraction (for example, those FDs/CFDs that involve a given attribute).

Relying on domain knowledge implies that the user is familiar with the dataset. However, if the dataset is new, then it requires exploration and extraction of patterns (as dependencies). In this case, existing mining tools are of little help. The plain list of dependencies is too bulky and too static to facilitate efficient information extraction. The problem here that data is of concern too: an analyst needs to work with both.

For example, suppose that the original data contains errors. The user has an idea that a given dependency holds, but a mining tool does not find it. In this case the user has to look into rows which contain violations to understand whether there is an actual dependency or not. Moreover, the user may want to try manual in-place repair to see how the data is affected by it and then, possibly, try to check the considered dependency again.

This is the first data cleaning sub-scenario which our approach aims to address. The next one is the post-repair validation. There are two types of data repair tools based on dependencies: automated and human-involved. Both of these types can result in unverified fixes, and even may introduce new errors~\cite{Ilyas:2019:DC:3310205}:

\say{It is often difficult, if not impossible, to guarantee the accuracy of any data repairing techniques without external verification by experts and trustworthy data sources.}

Therefore, post-validation is required. This step also implies exploring the data that has been repaired. 

The integrated workflow of both sub-scenarios is discussed in detail in Section~\ref{sec:Motivation:Workflow}.

\subsection{Integrity constraint checking}~\label{sec:Motivation:ICC}
Consider a process of verifying a specific integrity constraint~--- such a need may emerge within a newly formed database when no integrity constraints have been defined yet. Prerequisites for this use-case are the following:
\begin{enumerate}
    \item the user possesses certain knowledge regarding the data, or they are familiar with a subset of rules which describe patterns in data;
    \item use of heavy industrial-grade data management tools is unnecessary. For example, the data is already prepossessed and cleaned.  
\end{enumerate}

In this scenario we suggest to simply query a table for specific dependencies, which the user supposes to present in the data. Consider the following CFD on the Automobile table of a car dealership: for an automaker A1 and its model M2, engine type always defines engine volume. In other words, the dealership sells such A1M2 packages that for each engine type there is exactly one possible option for an engine volume. We can describe this rule in the notation of the new SQL predicate (Section~\ref{sec:SectionHolds}): HOLDS(``Automaker'', ``Model'', ``EngineType'' $\rightarrow$ ``EngineVol'' ON [``Automaker'' = ``A1''] and [``Model'' = ``M2'']). Validation of a single CFD is not a very time-consuming task like mining of all possible CFDs that present in a table. The result of the validation process is a subset of rows of the relation instance which agree on the CFD. There are three possible outcomes:
\begin{enumerate}
\item if the query returns all rows, then the hypothesis is true;
\item if the query returns almost all rows, then the user can invert the result (see NOT HOLDS predicate) to check which rows do not agree on the CFD. If the number of such rows is small, then table fixes can be performed manually from within a database.
\item if the query returns only a small fraction of rows (i.e., the CFD confidence is small), then there is a reason to reconsider the hypothesis or question the quality of the data.
\end{enumerate}

If the second scenario is realized, the user can conclude that the car dealership received wrong packages or a manager made a typo during a car registration procedure.

\subsection{Data source analysis}~\label{sec:Motivation:DSA}

Suppose a database table that is repeatedly populated from a number of sources. At the same time:

\begin{enumerate}
\item Data may come from different sources~--- web, user application, hardware sensors, etc;
\item Data may be of different quality and type. For example, it may be originally represented not only in a relational form, but in many others: nosql, graph, rdf etc. 
\end{enumerate}

After each data loading iteration, a need to analyze a specific data source may arise. 

For example, a data analyst may want to:
\begin{enumerate}
\item check whether a number of dependencies (conditional or regular ones) are still present;
\item know how current set of dependencies differs from its previous version.
\end{enumerate}

For this problem of repeated FD discovery, modern FD discovery algorithms allow to maintain an FD cover in a dynamic manner~\cite{DynFD}. However, such algorithms reside in standalone tools for data profiling (e.g., Metanome~\cite{Metanome}) and can not be run from inside a DBMS. 

In order to perform data analysis right in place without additional tools, we propose to implement the MINEFD operator and all the data structures necessary for incremental updating of the FDSet. This operator would allow a specialist to perform analysis conveniently, as shown in the following scenario:
\begin{enumerate}
\item a new portion of data arrives;
\item MINEFD is run to perform FDSet updating;
\item difference between current and previous FDSet versions is displayed (query FDSets with SELECTDEP clause).
\end{enumerate}

Now, it is possible to make conclusions regarding the newly-arrived data quality and its source; e.g. if the difference is substantial, then additional exploration of the data source may be required.

\begin{figure*}[ht]
    \centering
    \includegraphics[width = 0.9\linewidth]{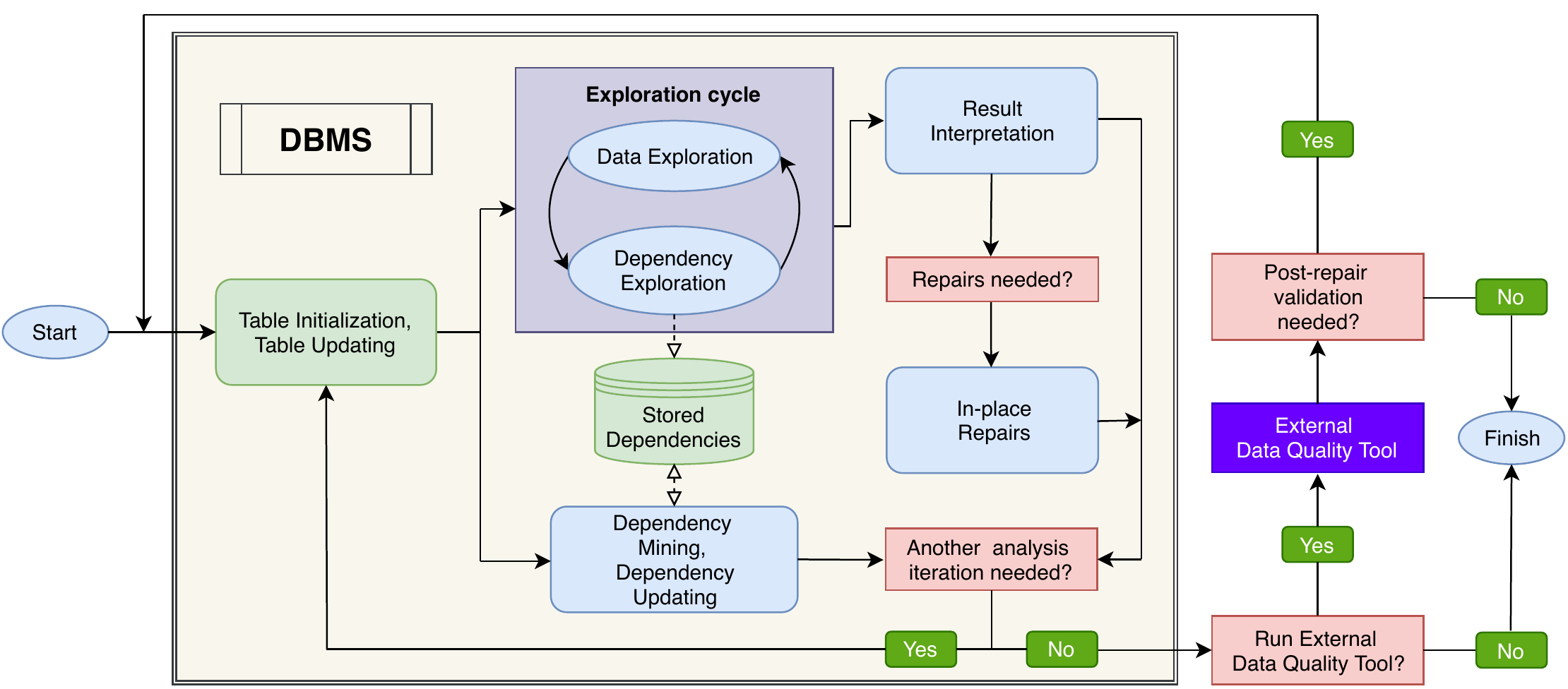}
    \caption{Workflow}
    \label{fig:Workflow}
\end{figure*}

\subsection{Dependency mining process specification}
\label{sec:Motivation:DMPS}

Mining dependencies for a medium-sized table is a very resource-consuming problem. Discovery becomes significantly more complicated on large tables, and sometimes it is not possible to mine all FDs, even restricting the length of the mined LHS. However, only a small fraction of the entire volume of discovered dependencies would be of interest for a data analyst.


Thus, in order to mine knowledge for big data applications, one needs to consider mining a subset of attributes of special interest. To the best of our knowledge, the usual discovery process runs on a stand-alone data profiling tool, which connects to a database or loads raw data. This tool allows to parameterize discovery process with maximum LHS length, number of CPU cores for parallel execution, error threshold, etc. However, no option to list attributes which participate in mined dependencies exists. Therefore, the only viable approach is to project an attribute subset onto a separate file or table, and then run the discovery process. The user will have to repeat this projection step each time they would like to specify a new subset of attributes. 

We propose to move the discovery process inside the DBMS, which results in a convenient way to specify an attribute subset for the following use-cases:
\begin{enumerate}
\item straightforward mining process employing all attributes is unfeasible due to table wideness or data peculiarities, so the user needs to remove some attributes from the search space;
\item data is already inside a DBMS and a fragmentation was performed. For such data located in different tables it becomes possible to perform reconstruction and list a subset of attributes.
\end{enumerate}

\subsection{Dependency exploration}
\label{sec:Motivation:DE}
Another problem regarding the large amount of mined dependencies is their exploration. Data profiling tools display discovered dependencies as plain text, lacking filtering options. According to an empirical study~\cite{Papenbrock:2016:HAF:2882903.2915203}, a table with only 27 attributes has 128K of functional dependencies, which makes manual browsing nearly impossible.

Moreover, a data analyst may deal with different kinds of FD sets:
\begin{itemize}
\item it could be a mixed set, comprised of both exact and approximate dependencies. Having a filtering process will allow to analyze the specific type of dependencies and form a number of integrity constraints.
\item the FD set could be augmented with dependencies that were manually specified. In this case, users can check if the current FDSet is still minimal as new dependencies arrived, or if they can be inferred from the existing FDSet.
\end{itemize}
Another problem that can arise is the consistency of representation of FDs a user will have to analyze. Since there could possibly be imported FDs alongside the mined ones, a user may eventually have to deal with a FD set comprised of groups of FDs from different sources. Each group represents FDs in its own way: for example, FD attributes represented by integers or their full names, or as a prefix tree view. Collecting such groups, transforming them into a unified representation and maintaining them inside a DBMS would spare a significant amount of time in the work of a data analyst.

Therefore, having a convenient way to navigate mined dependencies is an absolute necessity for performing in-database data analysis.

\begin{table*}[h]
    \centering
    \caption{IOWA Liquor (Projection).}
    \begin{tabular}{c|c|c|c|c|c|c|c|c|c|c|c|}
         \cline{2-12}
         $t_{id}$ & Date & Address & Zip & Category & CategoryName & Vendor & Pack & BtlVol & BtlSold & Sale & VolSold \\ \cline{2-12}
$t_1$ & 04-10 & COMM. AVE & 50533 & 62310 & BLACK RUM & 260 & 12 & 1000 & 3 & 50.82 & 0.79 \\ \cline{2-12}
$t_2$ & 12-17 & IOWA ST & 52001 & 12100 & WHISKIES & 115 & 12 & 1000 & 48 & 477.60 & 12.68 \\ \cline{2-12}
$t_3$ & 09-25 & ELM ST & 52001 & 62310 & BLACK RUM & 260 & 12 & 750 & 12 & 152.88 & 2.38 \\ \cline{2-12}
$t_4$ & 05-07 & 18TH ST & 50613 & 81600 & LIQUEUR & 259 & 12 & 750 & 3 & 22.68 & 0.59 \\ \cline{2-12}
$t_5$ & 03-12 & RIDGE RD & 52753 & 11200 &  BOURBON & 65 & 12 & 750 & 1 & 10.25 & 0.13 \\  \cline{2-12} 
$t_6$ & 07-02 & HWY 71 & 51333 & 12210 & SCOTCH & 65 & 12 & 750 & 12 & 82.70 & 1.82 \\ \cline{2-12}
$t_7$ & 01-15 & 8TH ST W & 50428 & 12210 & SCOTCH & 305 & 12 & 750 & 3 & 135.33 & 0.59 \\ \cline{2-12}
$t_8$ & 09-02 & HWY 71 & 51331 & 11200 & BOURBON & 65 & 12 & 750 & 12 & 179.40 & 2.38 \\ \cline{2-12}
$t_9$ & 04-10 & COMMERC. ST & 50533 & 32200 & VODKA & 260 & 12 & 750 & 4 & 65.32 & 0.79 \\ \cline{2-12}
$t_{10}$ & 08-14 & 2ND AVE & 50314 & 62310 & WHITE RUM & 125 & 6 & 750 & 1 & 26.66 & 0.20 \\ \cline{2-12}
    \end{tabular}
    \label{tab:example}
\end{table*}

\begin{table*}[h]
    \begin{minipage}{0.5\textwidth}
    \begin{lstlisting}[
        language=SQL,
        showspaces=false,
        basicstyle=\ttfamily,
        commentstyle=\color{gray},
        morekeywords={SELECTDEP, LENGTH},
        caption={Example query on IOWA Liquor FD set.},
        label={listing:FDMLExample},
        captionpos=b]
    SELECTDEP LHS -> RHS FROM FDSet
    WHERE 
    (LHS LIKE ({"Address", "Zip"} + 
    {"Address", "Category*"})
    AND LHS LIKE ("Sale", "Date"))
    OR (LHS LIKE ({"Vendor"})
    AND LHS LENGTH = 3
    AND RHS LIKE ("*Sold"))
    \end{lstlisting}
    \end{minipage}
    ~
    \begin{minipage}{0.5\textwidth}
    \begin{tabular}{|c|c|}
        \hline
        lhs & rhs \\ \hline
         Address, Zip & Sale  \\ \hline
         Address, Category & Sale  \\ \hline
         Address, CategoryName & Sale  \\ \hline
         Address, Zip & Date  \\ \hline
         Address, Category & Date  \\ \hline
         Address, CategoryName & Date  \\ \hline
         BtlVol, Category, Vendor & VolSold \\ \hline
         BtlVol, Category, Vendor & BtlSold \\ \hline
    \end{tabular}
    \centering
    \caption{Result table for Listing~\ref{listing:FDMLExample} query.}
    \label{tab:my_label}
    \end{minipage}
\end{table*}

\subsection{Workflow}~\label{sec:Motivation:Workflow}

In order to demonstrate the potential of our proposal, we present an example workflow in Figure~\ref{fig:Workflow}. It is a repeatable process involving data preparation along with data analysis. Such well-known data processing tasks are usually confined to stand-alone tools, each of which considers only its specific field of work. Contrary to this approach, we suggest to perform all FD-related data analysis and data cleaning within a single workflow inside a single DBMS.

For the sake of simplicity we suppose that a user works with a single table, but the workflow can be easily generalized for an arbitrary number of tables.

\textbf{Table initialize, table update.} During the very first step of initialization, a user populates the database with a new table (or a set of its fragments), performs schema mapping, etc. Since we consider the workflow as a repeating process, each subsequent iteration of the first phase would be updating the table according to the number of decisions regarding in-place data repairs.

Next, two steps are possible: exploration cycle and dependency mining.

\textbf{Dependency mining and dependency updating.} At this step the user populates the database dependency store via issuing commands that specify which dependencies to discover. The user is offered a mechanism that allows a flexible specification of dependencies to be mined: which attributes should participate and in which part (LHS or RHS), the length of sought dependencies and so on.

Alternatively, the user enters the \textbf{data exploration cycle}. This is the core of our proposal. The user can query dependencies associated with the table and use dependencies while querying the data itself. 

Having finished posing queries, the user can perform \textbf{result interpretation}, i.e. draw conclusions regarding the data and dependencies, summarizing extracted knowledge. If the user deems necessary to do so, they can perform \textbf{in-place data repairs}. 

After the repairs, the user may want to repeat the process from scratch to see the effects of repairs. If they consider data to be ready for an \textbf{external data quality tool}, they stop iterations and transfer control to the tool. Afterwards, if post-repair validation is needed, the user loads data back into the database and runs a few exploration cycles.

\section{Manipulating Data and Functional Dependencies}

\subsection{Querying Functional Dependency Collection}

A large table will likely result in a large collection of functional dependencies, which is cumbersome to browse manually. Indeed, the upper bound of the number of possible minimal, non-trivial dependencies is $\frac{N}{2}*2^N$~\cite{LiuDiscoverDependencies2012}, where $N$ is the number of attributes. Even if $0.1\%$ of potential dependencies hold, for a $20$-attribute table it will be around 20 thousand of functional dependencies to display.

In this section we present a special language designed to assist data scientists in the exploration of a collection of functional dependencies. We call it the Functional Dependency Manipulation Language (FDML). The full syntax of the language is presented in Listing~\ref{syntax:FDML}. We aimed to develop an SQL-friendly syntax for the following reasons:
\begin{inparaenum}[(1)]
\item it would be easier to understand the semantics of FDML for an SQL user
\item writing complicated queries of combinations of nested \texttt{SELECTDEP} statements with usual \texttt{SELECT} on rows of a table becomes an intuitive task.
\end{inparaenum}

 FDML contains clauses known to every user of SQL~--- \texttt{WHERE} and \texttt{LIKE}, which are designed to help specify conditions on objects we are working with. 
 Since the language is designed for dependency manipulation, we need a powerful tool for both left- and right-hand side specification. 
 Thus, the core part of FDML syntax is comprised of the \texttt{LhsCondition} and \texttt{Subset} clauses with their ability to eloquently describe attribute sets. 
 Both clauses represent a set of rules to express operations on subsets of attributes.

 Consider an example query: ``check if FD with the following conditions is present in a table:  Lhs is [any attribute except for A] OR [A and B] and Rhs is C''. 
 Obviously, to manage the example-like queries we need some kind of a regular expressions grammar. 
 It could be also very useful to enumerate sets of attributes which have common patterns in their naming. For example, for a set of attributes \{``data1'', ``data2'', \ldots\} the regexp in the notation of \texttt{Subset} would be \{``date*''\}, which means we consider any attribute that represents information on data as a part of a dependency.
 
LHS specification is controlled by \texttt{LhsCondition}, which allows to specify complicated conditions on subsets which may be presented in a left-hand side of a dependency. In Listing~\ref{listing:FDMLExample} the first part of the query (before OR) we can read as follows: ``find all FDs whose LHS contain attribute sets [(Address and Zip) or ((Address, Category) or (Address, CategoryName))] and RHS must contain (Sale and Date)''. The result of that part are the first six rows presented in Table~\ref{tab:my_label}.

FDML provides users with ability to specify two more conditions on dependencies: 

\begin{itemize}
    \item \texttt{ERROR} threshold of relaxation on a dependency. When working with AFDs, for each of them we can also store the error measure $e (X \rightarrow Y, r)$ as an individual attribute of \texttt{FDSet} table. 
    \item \texttt{LENGTH} of LHS. The usage of the predicate is presented in the second part of Listing~\ref{listing:FDMLExample} query. In that part we consider only dependencies with the following properties: 
    \begin{inparaenum}[(1)]
\item LHS must contain the Vendor attribute
\item LHS length is equal to three (i.e., dependencies where Vendor determines RHS with no more or less than two other attributes with free of choice)
\item RHS must contain attributes that match the pattern ``*Sold''.
\end{inparaenum}
\end{itemize}

We have demonstrated an initial draft of FDML that can be further extended with such inference rules as: 1) returning all paths that represent transitive dependencies from X to Y, 2) calculate a closure of a set of attributes, 3) check if the specified FD can be derived from the current FD set.

\subsection{Extending SELECT Clause to Support FD-related Filters}\label{sec:SectionHolds}


In this section we discuss an extension to the SELECT clause that allows to query data using FD-related predicates. Let us review the proposed statements and their respective use-cases.

\subsubsection{HOLDS}

The first proposed extension aims to allow the user to browse all rows that conform to a specified functional dependency. For this purpose, we introduce a novel predicate \texttt{HOLDS} that selects all rows which have equal RHS parts for the same, shared LHS. 

Note that since we do not know which one of the records contains the ``right'' value, the semantics of this predicate are not straightforward. There are several possible approaches to defining it: 
\begin{enumerate}
    \item Do not include any row that has a contradicting RHS into the result.
    \item Include only a single row: first, random or according to some criteria (e.g. median value of RHS).
\end{enumerate}

We strongly believe that the first approach should be used for the sake of usability. Indeed, it would be more flexible to allow the user to select the ``right'' value as they deem relevant, using the \texttt{NOT HOLDS} predicate and existing SELECT sub-clauses.

Furthermore, we would like to make it possible to specify other conditions while querying data with functional dependency predicates:

\begin{lstlisting}[
        language=SQL,
        showspaces=false,
        basicstyle=\ttfamily,
        commentstyle=\color{gray},
        morekeywords={HOLDS, WITH},
        caption={Combining HOLDS and other predicates.},
        captionpos=b,
        label={ex:error}]
SELECT "Category", "BtlVol", 
"CategoryName" FROM IOWA
WHERE HOLDS ("Category", "BtlVol" -> 
"CategoryName") AND "BtlVol" >= 750 
\end{lstlisting}

However, while designing this operator, we have identified a problem: \texttt{HOLDS} (and other FD-related predicates) does not commute with regular predicates. In other words, the results of \texttt{WHERE HOLDS ("Category", "BtlVol" -> \\ "CategoryName") AND "BtlVol" >= 750} depend on the order of predicate evaluation. To resolve this ambiguity, we have decided that all FD-related predicates are to be evaluated first. In case when a user needs to restrict rows before evaluating such predicates, it is possible to parameterize them as shown in the next example.

\begin{table}[H]
    \begin{minipage}{0.5\textwidth}
    \centering
    \begin{lstlisting}[
        language=SQL,
        showspaces=false,
        basicstyle=\ttfamily,
        commentstyle=\color{gray},
        morekeywords={HOLDS},
        caption={Fetch all rows holding the specified FD on attributes Category, BtlVol and CategoryName with additional conditions. Result: 5-8-th rows in IOWA Liquor.}, 
        captionpos=b,
        label={listing:filtering}]
SELECT "Category", "BtlVol", 
"CategoryName" FROM IOWA
WHERE HOLDS ("Category", "BtlVol" -> 
"CategoryName" ON ["BtlVol" >= 750] 
AND (["Category" = 11200] 
OR ["CategoryName" = "SCOTCH"))
	\end{lstlisting}
    \end{minipage}

    \vspace{-0.2cm}
    \begin{minipage}{0.5\textwidth}
    \centering
    \caption*{Tableau $T_2$}
    \begin{tabular}{|l|l||l|}
        \hline
        \textbf{Category} & \textbf{BtlVol}  & \textbf{CategoryName} \\ \hline
        11200 & 750 &  - \\ \hline
        - 	  & 750 &  SCOTCH \\ \hline
        11200 & 1000 &  - \\ \hline
        -	  & 1000 &  SCOTCH \\ \hline
        \end{tabular}
    \label{table:CFDTableau2}
    \end{minipage}
\end{table}



Finally, one can notice that any complex conditions on attributes can be represented with a CFD pattern tableau. For example, for query shown in Listing~\ref{listing:filtering} its pattern tableau is $T_2$.

The proposed operator can be expressed by the means of existing SQL operators as shown in Listing~\ref{listing:equivalentHolds}. However,
\begin{itemize}
    \item the equivalent query is rather large and confusing. An increase of the LHS size will deepen these issues.
    \item the equivalent query is also expensive: it performs a join and a subquery (although a non-correlated one) with aggregation. At the same time, modern FD discovery algorithms rely on a rather computationally cheap operation of partition intersection.
\end{itemize}

\begin{lstlisting}[
        language=SQL,
        showspaces=false,
        basicstyle=\ttfamily,
        commentstyle=\color{gray},
        morekeywords={HOLDS, WITH},
        caption={SQL equivalent of HOLDS query presented in Listing~\ref{listing:filtering}.},
        captionpos=b,
        label={listing:equivalentHolds}]
WITH fdtemp AS (SELECT "Category", "BtlVol" 
FROM IOWA WHERE ("BtlVol" >= 750 
AND "Category" = 11200) 
OR "CategoryName" = 'SCOTCH'
GROUP BY "Category", "BtlVol"
HAVING (COUNT (DISTINCT "CategoryName")) = 1)

SELECT "Category", "BtlVol", "CategoryName" 
FROM IOWA JOIN fdtemp
ON fdtemp."Category" = "Category" 
AND fdtemp."BtlVol" = "BtlVol"
\end{lstlisting}

\subsubsection{Approximate HOLDS}

Since FDs rarely hold on real data, we propose to extend \texttt{HOLDS} with the capability to include approximate FDs. For this, we introduce the \texttt{ERROR} parameter, which is in fact the error threshold $e_{max}$ for AFDs we defined in Section~\ref{sec:Basics}. It is used as follows:

\begin{lstlisting}[
        language=SQL,
        showspaces=false,
        basicstyle=\ttfamily,
        commentstyle=\color{gray},
        morekeywords={HOLDS},
        caption={Fetch rows that conform to the specified FD (Category $\rightarrow$ CategoryName). Result: 
        Rows 2, 4-9 in IOWA Liquor.},
        captionpos=b,
        label={ex:error}]
SELECT "Category", "CategoryName"
FROM IOWA WHERE 
HOLDS ("Category" -> "CategoryName", 
ERROR = 0.05)
\end{lstlisting}

To fetch a result for such a query, we first need to calculate the error measure $e(Category \rightarrow CategoryName, IOWA)$ according to Definition~\ref{def:ErrorMeasure}.
We determine pairs of tuples (and their inverses) which violate the specified dependency: $(t_1, t_{10})$, $(t_3, t_{10})$, $(t_{10}, t_1)$, $(t_{10}, t_3)$. Thus, $e(Category \rightarrow \\ CategoryName, IOWA) = \frac{4}{10^2 - 10} \approx 0.04 \leq 0.05$. Since the error measure value on the AFD is less than the given threshold, we can return all rows except for the ones which disagree on the CategoryName attribute. The result is rows 2 and 4-9 in IOWA Liquor.

To express this query using existing SQL statements a considerable effort would be required:

\begin{enumerate}
    \item First of all, a numerator of $e(Category \rightarrow \\ CategoryName, IOWA)$ should be computed:

\begin{lstlisting}[
        language=SQL,
        showspaces=false,
        basicstyle=\ttfamily,
        commentstyle=\color{gray},
        morekeywords={HOLDS},
        caption={Computing numerator.},
        captionpos=b,
        label={ex:numeratorerror}]
        
CREATE AGGREGATE mul(bigint) 
    ( SFUNC = int8mul, STYPE=bigint );

WITH fdtemp AS (SELECT Category as cat, 
(COUNT (Categoryname)) AS cn FROM IOWA
GROUP BY Category, categoryname)

SELECT cat, mul(cn) * 2 as numerator 
FROM fdtemp
GROUP BY cat HAVING COUNT(*) > 1
\end{lstlisting}

    \item Next, computing denominator requires extracting the total number of rows:
    
\begin{lstlisting}[
        language=SQL,
        showspaces=false,
        basicstyle=\ttfamily,
        commentstyle=\color{gray},
        morekeywords={HOLDS},
        caption={Computing denominator.},
        captionpos=b,
        label={ex:denominatorerror}]
        
SELECT COUNT(*) AS row_num FROM IOWA
\end{lstlisting}    

    \item Finally, we have to run the following check:

\begin{lstlisting}[
        language=SQL,
        showspaces=false,
        basicstyle=\ttfamily,
        commentstyle=\color{gray},
        morekeywords={HOLDS},
        caption={Check.},
        captionpos=b,
        label={ex:checkerror}]
        
CASE WHEN fdtemp.numerator / 
(row_num^2 - row_num) <= 0.05:
THEN *run query itself*
ELSE *perform nothing*

\end{lstlisting}

    \item The query itself, mentioned in the previous listing is essentially query from Listing~\ref{ex:error} but without the ERROR threshold. How to rewrite such queries using SQL means was discussed in the HOLDS subsection.
    
\end{enumerate}

\subsubsection{NOT HOLDS}

Next, it is possible to invert the selection predicate in order to obtain all rows where the specified FD does not hold. Again, in this case the same ambiguity problem arises. We resolve it in the same manner: via returning all candidate rows. The example is as follows:

\begin{lstlisting}[
        language=SQL,
        showspaces=false,
        basicstyle=\ttfamily,
        commentstyle=\color{gray},
        morekeywords={HOLDS},
        caption={Fetch all rows violating the Address $\rightarrow$ Zip FD. Result: 
        6 and 8 rows in IOWA Liquor.},
        captionpos=b,
        label={ex:notholds}]
SELECT * FROM IOWA
WHERE NOT HOLDS ("Address" -> "Zip")
\end{lstlisting}

The SQL equivalent of this operator is presented below.

\begin{lstlisting}[
        language=SQL,
        showspaces=false,
        basicstyle=\ttfamily,
        commentstyle=\color{gray},
        morekeywords={HOLDS},
        caption={SQL equivalent of NOT HOLDS query presented in Listing~\ref{ex:notholds}.},
        captionpos=b,
        label={listing:equivalentNotHolds}]

WITH fdtemp AS 
(SELECT Address AS Addr FROM IOWA
GROUP BY Address
HAVING (COUNT (DISTINCT ZIP)) > 1)

SELECT Address, Zip FROM IOWA
JOIN fdtemp ON
Address = fdtemp.addr
\end{lstlisting}

In the first part of this query we select all Zips which contain several different values for the same address. In the second we join them with the original table to extract rows that have the same Address, but different in Zip. Again, there are the same issues as with simulating HOLDS clause: costly operations and unclear semantics.






\subsubsection{VIOLATES}

Next extension is the \texttt{VIOLATES} clause. Its intended scenario is a user surmising that there is a functional dependency in data, but there are errors in some of the LHS attributes. The goal is to select such rows for further inspection or correction. Consider the following query:

\begin{lstlisting}[
        language=SQL,
        showspaces=false,
        basicstyle=\ttfamily,
        commentstyle=\color{gray},
        morekeywords={VIOLATES},
        caption={Fetch all rows violating the specified FD on attribute Address. Result: rows 1, 9 in IOWA Liquor.},
        captionpos=b,
        label={listing:violates}]
SELECT * FROM IOWA WHERE "Address" 
VIOLATES ("Address", "Vendor" -> "Zip", 
ERROR <= threshold)
\end{lstlisting}

Here, the user suspects an error in $Address$, with respect to the $Address, Vendor \rightarrow Zip$ FD. To determine possible inaccuracies, the following steps need to be performed: 
\begin{inparaenum}[(1)]
\item group values by Zip and Vendor attributes,
\item measure a distance between values on attribute Address inside each group,
\item the result would be all rows with distance less or equal to a threshold value.
\end{inparaenum}
Indeed, in our example rows $t_1$ and $t_9$ are likely to contain inaccuracies and require correction.

Implementing VIOLATES using SQL clauses will require creation of a stored procedure that computes the distance between records. For example, for query presented in Listing~\ref{listing:violates} it is necessary to create a stored procedure that computes a metric on strings, since there is Address attribute that has Varchar data type. This procedure should be invoked for all records that were returned by HOLDS (``Address'', ``Vendor''$\rightarrow$ ``Zip'') subquery. 
This metric should be used to check all records with the same Vendor and Zip but with different Address.

\subsubsection{DEPENDENT}

The next clause is a projection which keeps attributes that depend on the specified collection of attributes. It can also feature the \texttt{ERROR} constraint.

\begin{lstlisting}[
        language=SQL,
        showspaces=false,
        basicstyle=\ttfamily,
        commentstyle=\color{gray},
        morekeywords={DEPENDENT},
        caption={Dependency projection. Result is all rows of the original table, but only for the attributes Date, Sale, CategoryName, VolSold, Category.},
        captionpos=b]
SELECT DEPENDENT (["Zip", "Address"]) 
FROM IOWA
\end{lstlisting}

Expressing this operator in terms of classic SQL is much harder since the constraints are applied not to values, but to attributes. Therefore, it would likely result in enumerating different combinations of attributes within HAVING COUNT operator.


\subsubsection{Concluding Remarks}

We have presented a number of operators that will allow user to manipulate data using dependencies. As we can see, all of these operators can be expressed by classic SQL statements. However, having a special statement for HOLDS (exact and especially approximate), VIOLATES, and DEPENDENT can considerably simplify query syntax.

\section{Mining and Storing Discovered Functional Dependencies}

In the previous section, we have described a number of clauses that allow to query a collection of FDs. However, we left out an important question: where do these FDs come from? There are two possibilities: mining functional dependencies either online or offline. In the first case, FDs are mined on-the-fly and then filtered according to query predicates. However, this approach is rather costly in terms of required computational resources. Contemporary FD mining algorithms may work for hours or days, even on smaller tables~\cite{Pyro}. Note that these numbers were obtained in an environment where all data was residing in memory. Therefore, this approach is not suitable at all for tables that are too big to fit into memory.

Hence, the only practically viable approach is to pre-mine FDs and to store them for later use. We propose to keep a special data structure for each table: an FD-table. It is designed to store associated FDs. However, it is also intended for acting as a mediator between user queries and underlying data. For this, we propose a special statement presented in Listing~\ref{statement:minefd}.

This will allow to fine-tune the number and type of mined dependencies. The run times presented above are specified for the mining of all non-trivial and minimal functional dependencies that are present in a given table. The majority of mining algorithms compute FDs incrementally by ``growing'' the size of LHS. This process can be viewed as lattice traversal~\cite{TANE, Papenbrock:2016:HAF:2882903.2915203}, where each level corresponds to an individual LHS size. The process can be interrupted on a given level, allowing to select FDs with LHSes of specified lengths.

Interrupting the mining algorithm can save not only time, but also space required to store FDs. Let us recall that the number of potential FDs grows fast and may become huge even for tables spanning a few dozens of attributes. Storing all FDs will require significant space which may easily exceed that required to store the original table. At the same time, one may say that for an analyst, the usefulness of discovered FDs decreases with the increasing size of its LHS. Therefore, it may be viable to specify the maximum length of the LHS. Furthermore, it is possible to specify other constraints such as names of eligible attributes in order to further narrow down the search space.
\section{Problems and Challenges}

This proposal arises many questions and opens up a multitude of research venues. Some of them are:

\begin{enumerate}
    \item Query processing. The Volcano model should be extended with a number of new operators. Although the majority of them are simple projections and selections, they still require work. Furthermore, existing complex operators like joins or aggregation require critical reflection in order to understand how they affect FDs of base tables.
    \item Query optimization. Novel operators will require updating optimizers: novel algebraic equivalences and novel cost models will appear.
    \item Choosing to stay within the relational model or not. Designing novel operators will require a compromise between the features and strictness of the proposed operators.
    \item Indexing FD collections. Having millions of FDs and an expressive query language for them, there will be a need for fast answers. Therefore, development of novel indexing methods will be relevant.
    \item Data updates. Data updates will require special treatment of the FD-table in order to keep it up-to-date. Therefore, it will require further development of dynamic FD discovery algorithms like DYNFD~\cite{DynFD}.
    \item Operator semantics and uncertainty. In Section~\ref{sec:SectionHolds} we have discussed several cases where uncertainty may arise. 
    \item Focused FD discovery. The \texttt{MINEFD} operator narrows down the search space in order to increase the efficiency of mining. It allows not only to restrict the length of mined FDs, but other properties as well. Development of efficient FD discovery algorithms that search in a given neighborhood becomes an important research problem.
    \item FD mining for disk-based datasets. So far, all FD discovery algorithms assumed that all data fits into memory, which is not true for real data. Developing specialized FD mining algorithms will help solve this problem.
    \item General FD-table handling, handling of several FD-tables. There are multiple questions. How do we handle the FD-table? Should it be transparent to the user or not? Should we allow several different FD-tables for a single table? If yes, how do we organize the interactions?
    \item Comparing workflows that employ dependencies to other approaches for extracting knowledge, e.g. association rules.
\end{enumerate}

\section{Conclusion}

In this paper we have described our proposal to make functional dependencies first-class citizens in RDBMSes in order to provide analysts with a tool that allows to use them during data exploration. We have described a language for querying a collection of FDs and SQL extension that allows to employ FD-related conditions during SELECTs. Besides, we have presented some thoughts related to mining and storage of mined FDs inside a RDBMS. Finally, we have discussed  implementation challenges and outlined prospective research venues.


\bibliographystyle{abbrv}
\bibliography{vldb_sample}  



\begin{appendix}


\small
\begin{lstlisting}[
        language=SQL,
        showspaces=false,
        basicstyle=\ttfamily,
        commentstyle=\color{gray},
        morekeywords={SELECTDEP},
        caption={Functional Dependency Manipulation Language (FDML) syntax.},
        label={syntax:FDML},
        captionpos=b]
SELECTDEP [ * | <Lhs> '->' <Rhs> | ] 
FROM <FDSet>
[WHERE <Condition>]

<Condition> := 
    <RhsCondition> 
    | <LhsCondition> 
    | 'ERROR' <real>
<RhsCondition> := <Rhs> 'LIKE' <Subset>
<LhsCondition> := 
    <Lhs> 'LIKE' <LhsConstruct> 
    | <Lhs> 'LENGTH' <operator> <uint>

<Subset> := 
    '{' <RegExpList> '}'
    | * 
    | '(' <Subset> '+' <Subset> ')'
    | '(' <Subset> '-' <Subset> ')'

<RegExpList> := 
    <RegExp> 
    [ ',' <RegExpList>]

<LhsConstruct> := 
    '[ AND' <Subset> ']'
    | '[ OR' <Subset> ']'
    | *
    | '(' <LhsConstruct> '+' <LhsConstruct> ')'
    | '(' <LhsConstruct> '-' <LhsConstruct> ')'
\end{lstlisting}

\begin{lstlisting}[
        language=SQL,
        showspaces=false,
        basicstyle=\ttfamily,
        commentstyle=\color{gray},
        morekeywords={MINEFD, SELECTDEP},
        caption={MINEFD operator syntax.},
        label={statement:minefd},
        captionpos=b]
MINEFD <FDResultSet> AS
SELECT <Lhs> '->' <Rhs> [',' ERROR]
[WHERE <Condition>] 
FROM <TableName> ['ERROR' <real>]

<Condition> :=
    <Lhs> 'LIKE' <Subset>
    | <Rhs> 'LIKE' <Subset>
    | <Lhs> 'LENGTH' <operator> <uint>
\end{lstlisting}

\end{appendix}

\balance

\end{document}